# SURFACE DENSITY EFFECTS IN QUENCHING: CAUSE OR EFFECT?

Simon J. Lilly and C. Marcella Carollo

Institute for Astronomy, Department of Physics, ETH Zurich, 8093 Zurich, Switzerland

ABSTRACT

There are very strong observed correlations between the specific star-formation rates (sSFR) of galaxies and their mean surface mass densities, $\Sigma$, as well as other aspects of their internal structure. These strong correlations have often been taken to argue that the internal structure of a galaxy must play a major physical role, directly or indirectly, in the control of star-formation. In this paper we show by means of a very simple toy model that these correlations can arise naturally without any such physical role once the observed evolution of the size-mass relation for star-forming galaxies is taken into account. In particular, the model reproduces the sharp threshold in $\Sigma$ between galaxies that are star-forming and those that are quenched, and the evolution of this threshold with redshift. Similarly, it produces iso-quenched-fraction contours in the $f_Q(m,R_e)$ plane that are almost exactly parallel to lines of constant $\Sigma$ for centrals and shallower for satellites. It does so without any dependence on quenching on size or $\Sigma$, and without invoking any differences between centrals and satellites, beyond the different mass-dependences of their quenching laws. The toy-model also reproduces several other observations, including the sSFR gradients within galaxies and the appearance of inside-out build-up of passive galaxies. Finally, it is shown that curvature in the Main Sequence sSFR-mass relation can produce curvature in the apparent B/T ratios with mass. Our analysis therefore suggests that many of the strong correlations that are observed between galaxy structure and sSFR may well be a consequence of things unrelated to quenching and should not be taken as evidence of the physical processes that drive quenching.

*Subject headings*: galaxies: evolution – galaxies: structure – galaxies: bulges – galaxies: high redshift

1.  INTRODUCTION

The evolving galaxy population displays a number of empirical relations between the parameters that describe individual galaxies. Study of these global relations within the population may give us insights into the evolution of individual galaxies, partially overcoming the well-known limitation that we can only observe individual galaxies at a single snapshot of their development. Nevertheless, establishing the causality that lies behind these empirical relations, i.e. whether they are the result of a direct physical link, or whether they simply reflect a common dependence on other quantities, is hard to establish. The processes controlling star-formation in galaxies are a case in point.

Galaxies may be broadly divided into two categories: in one, the overall star-formation rate is quite tightly coupled to the existing stellar mass of the galaxy, producing a so-called "Main Sequence" (MS) in which the specific star-formation rate (sSFR) varies only weakly with mass, exhibiting a scatter around the mean sSFR-mass relation of around 0.3 dex. In the second, the sSFR is evidently suppressed by one or two orders of magnitude, or more. These galaxies form the so-called "Red Sequence" of passively evolving systems. Understanding the process or processes that cause MS star-forming galaxies to effectively cease forming stars at some point in their development is a major goal of extragalactic astrophysics. In this paper we will refer to this process as "quenching".

A small minority of galaxies have sSFR significantly above that of the MS. At least locally, these star-burst systems are predominantly associated with mergers of galaxies and this may well be true at all redshifts. There is evidence that the fraction of these "outliers" stays more or less constant with redshift, at least back to $z \sim 2$. We will not consider these objects further in this paper.

Application of straightforward continuity equations to the evolving mass functions $\phi(m)$ of star-forming and passive galaxies has produced a good phenomenological description of the outcome of quenching over cosmic time that will be used in this paper. The important observation that the mass function of star-forming galaxies has exhibited a constant characteristic Schechter mass M* (e.g. Ilbert et al 2013) and a more or less constant Schechter faint end slope $\alpha$ since $z \sim 4$ (see Peng et al 2014 and references therein) places strong constraints on the forms of quenching, whether expressed as a quenching rate $\eta(m,z)$ or as the survival probability to reach a given mass $P(m)$. Peng et al (2010) defined two quenching channels that are consistent with these constraints. The first is strongly mass-dependent and is called "mass-quenching". It is responsible for maintaining the constant M* of the star-forming population and produces a population of quenched passive galaxies that has this same M* as the star-forming population, but an $\alpha$ that differs by approximately unity. The quite different shapes of these two mass functions produces a fraction of quenched objects that increases steeply with mass. The second process is independent of mass and is called "environment quenching". It produces a second component of the mass function of passive galaxies with the same M* and the same $\alpha$ as the star-forming population. Peng et al (2012) showed that, while mass-quenching acts on all galaxies, the effects of environment-quenching are mostly visible in the population of satellite galaxies (galaxies that are in the halo of a larger central galaxy), since only those relatively few centrals in large groups show any environmental effects (Knobel et al 2015). It may be that mass- and environment-quenching are closely related despite their apparently quite different mass-dependences (see Carollo et al 2016, Knobel et al 2015 for discussion)

A strong increase in $\phi^*$ of galaxies with cosmic epoch is observed (e.g Ilbert et al 2013) and is almost unavoidable in the continuity analysis of Peng et al (2010) if M* is constant and the slope of the star-forming mass function $\alpha_{SF} < -1$. This evolution in $\phi^*$ emphasizes that the build-up of the passive population takes place over an extended period of time as newly quenched galaxies are added to the previously quenched population. This in turn makes clear that changes in the mean properties of the members of the passive population can be due to the addition of new members (with different properties) as well as to possible evolutionary changes in the properties of existing individual members. In this paper, we will refer to the former as *progenitor effects*. Progenitor effects have been extensively discussed in the context of the observed size evolution of the passive population in Carollo et al (2013).

Despite this progress in the phenomenology, the physical processes causing these quenching processes remain uncertain and consequently much debated in the literature. In particular, the relative role of the external halo environment and of internal processes operating within the galaxies is not known. Temperature and cooling effects in the halo have long been suspected in controlling the formation of stars in halos (Rees & Ostriker, 1977, White & Rees 1978, Blumenthal et al 1984). On the other hand, as reviewed below, strong correlations between the sSFR and gross structural properties of galaxies have also long been known, suggesting that internal processes within galaxies are responsible. These could be direct, if galaxy structure controls star-formation through, for example, disk stability (Martig et al 2009, Genzel et al 2014), or indirect, if galaxy structure is linked to, for example, the properties of a central black-hole, since energy injection from black hole accretion has been invoked as a mechanism both to expel the interstellar medium from galaxies (Springel et al 2005, Hopkins et al 2006) and to heat the gas in the surrounding halo (Croton et al 2006).

The idea that the stellar surface density of a galaxy might be playing a dominant role in the quenching of star-formation in a galaxy has recently gained a lot of attention. In an early SDSS analysis, Kauffmann et al (2003) pointed out the existence of a transition surface mass density. Defining the mean surface mass density $\Sigma_e$ within the half-light radius $R_e$, Kauffmann et al showed that there was a threshold at $\Sigma_e = 10^{8.5}$ $M_\odot \text{kpc}^{-2}$. Above this threshold, most galaxies have a large $D_n(4000)$ index indicative of a passive stellar population, while below it the low $D_n(4000)$ values typical of star-forming galaxies dominate. This is more explicitly seen in Brinchmann et al (2004) who computed the mean log sSFR as a function of $\Sigma_e$. Franx et al (2008) extended this analysis to very high redshifts, and showed that the threshold surface mass density steadily increases with redshift. The links between sSFR and galaxy structural parameters are also seen in terms of velocity dispersion (see e.g. Smith, Lucey and Hudson 2009, Wake, Franx and van Dokkum 2012) and light profiles, for example Sersic indices (see e.g. Blanton et al 2003, Wuyts et al 2011 amongst many others). Since these different parameterizations of structure are all inter-related it is unclear whether any of them is more important that the others. In this paper we will focus primarily on stellar mass density as an easily modellable quantity, but we would expect the broad conclusions to apply to other parameterizations also.

Omand et al (2014) have recently examined the fraction of galaxies that are quenched $f_Q$ as a function of stellar mass and half-light radius $R_e$, for both central and satellite galaxies, in the SDSS. Figure 9 in their paper again makes clear that the quenched fraction is indeed much better predicted by the mean surface density of the galaxy within $R_e$, i.e. $\Sigma_e = m/2\pi R_e^2$, than by its stellar mass. Indeed, the contours of iso-quenched-fraction on the left hand panel of their Figure 9 (for central galaxies) are strikingly parallel to the lines of equal surface mass density $\Sigma_e$. These authors therefore regarded it as a self-evident fact that quenching could not depend on galaxy stellar mass alone, as in the Peng et al (2010) formalism. They furthermore concluded, because the iso-quenched-fraction contours for satellite galaxies had a different slope, that satellite quenching must be accompanied by structural change. Similarly, Woo et al (2015) have plotted the quenched fraction of central galaxies as a function of halo mass and the stellar surface mass density within the central 1 kpc, showing that there was an abrupt change in $f_Q$ at a $\Sigma_{1\text{kpc}} \sim 10^{9.5} M_\odot \text{kpc}^{-2}$ inferring that processes within galaxies as well as in haloes were playing a role in quenching star-formation.

Several theoretical studies have explored ways in which the internal structure of galaxies could severely suppress star-formation in so-called morphological- or gravitational-quenching. Many of these have been based on considerations of disk stability via the Toomre $Q$ parameter (see e.g. Martig et al 2009, Genzel et al 2014).

However, while the observational link between structure and the quenched-state of a galaxy is very well-established, it is in our view not clear whether this reflects a direct causal connection or whether it in fact arises from processes that do not involve surface mass density (or velocity dispersion, etc) at all. In other words, do the evident strong links with structure reflect the "cause" of the quenching, or a "side effect" of something else completely?

The point of this paper is to show that the strong (apparent) connections between surface mass density and the quenching of galaxies follow from a single well-established completely independent fact about the evolving population of galaxies: the half-light radius $R_e$ of star-forming galaxies at a given instantaneous mass decreases with redshift, the size going as roughly $R_e \propto (1+z)^{-1}$. We will argue in this paper that the strong empirical sSFR-$\Sigma_e$ relation *cannot* therefore be taken to imply any causal physical link between $\Sigma$ and quenching.

With modern deep surveys of the distant galaxy population, it has become relatively straightforward to measure the observed $R_e$ of both star-forming and passive galaxies. While much of the focus has been on the latter because of interest in the apparent size-growth of the passive population with time, the equivalent behavior of the star-forming population is equally clear. For example, Buitrago et al (2008) observed growth (for $m > 10^{11}$ M$_\odot$ "disk-like" galaxies) of $R_e \propto (1+z)^{-0.8}$ to $z \sim 3$. More recently, amongst many studies, Newman et al (2012) show $(1+z)^{-1}$ over $0.5 < z < 2.5$ while Mosleh et al (2014) have observed $(1+z)^{-1}$ over $1 < z < 6$. There is still uncertainty in the precise exponent, and in the redshift range of its validity, but we will take $(1+z)^{-1}$ as a simple starting point for our analysis. The sensitivity to this assumption will be explored below.

It should be noted that the "virial radius" of dark matter haloes will also scale, at fixed mass, as $(1+z)^{-1}$. This is by construction, since the "virial density" used to set the virial radius is defined to be a constant multiple of the mean cosmic density. Nevertheless, having the visible parts of galaxies follow the same scaling is quite reasonable if the collapse factor of the star-forming baryons relative to the halo virial radius is more or less constant. Quite apart from any theoretical motivations along these lines (Mo et al 1998), the observed scaling for galaxies of $(1+z)^{-1}$ is phenomenological entirely reasonable.

At any epoch, the quenched galaxies will always be smaller and have higher $\Sigma$ than their star-forming counterparts (of the same mass), simply because the star-forming progenitors of the quenched galaxies would have been smaller at the earlier epoch at which they ceased forming stars. Likewise, at a given mass, we would expect a recently quenched galaxy to be larger, and thus have lower $\Sigma$, than one that quenched long ago. A given galaxy that is continuing to form stars will also have been smaller at earlier times, both because of the $(1+z)$ factor and because it was, at earlier times, also of lower mass. This of course produces an inside-out growth of galaxies in which the galaxy continually increases its half-mass and half-light sizes as it also adds mass.

The $(1+z)$ dependence of the scale size with redshift (at fixed mass) has the interesting feature that, combined with a roughly $m^{1/3}$ scaling with mass (see Equation 3 later in the paper), means that there will be a broad connection between the average physical density of stars in a star-forming galaxy and the mean cosmic density at that epoch.

These ideas then open up the possibility of explaining the striking observed correlations between the mean surface mass density $\Sigma$ and the star-formation state of a galaxy in terms of the link between density and epoch, rather than in terms of any physical connection between density and the control of star-formation. In other words, does the higher density simply reflect the earlier epoch at which these stars formed rather than a causal link between density and the control of star-formation?

The aim of this paper is to explore the implications of the observed size of star-forming galaxies on the build-up of stellar mass in galaxies. In particular, we will explore the extent to which it can explain the observed trends of star-formation history with $\Sigma$ *without invoking any physical effects directly related to size or $\Sigma$*. To do this we construct a simple toy model for the evolution of individual galaxies and use this to construct a model population of galaxies. As with any such toy-model, the intention is not to try to follow all relevant physical processes, nor to provide a precise quantitative comparison with observations, but rather to explore a very simple scenario in which the consequences can be very clearly followed. We focus on surface mass-density as it is relatively straightforward to model, but the ideas will equally well apply to central velocity dispersion etc., via the virial theorem.

The layout of the paper is as follows. In Section 2, we construct our toy model for exploring the build-up of the stellar populations in a population of model galaxies. This is anchored on just two empirical facts, namely the sSFR(*m,z*) and R$_e$(*m,z*) of star-forming Main Sequence galaxies. To these are added the empirical mass-dependent quenching laws of Peng et al (2010, 2012). We then validate this population model by checking the resulting overall mass functions. In Section 3, we then study a number of outputs of the model, demonstrating that the model successfully reproduces many of the sSFR-Σ effects discussed above, as well as several other observed phenomena, without invoking any physical processes that are causally linked to Σ. In Section 4 we offer some brief further discussion and then in Section 5 present a summary of our main conclusions.

The model is computed using the redshift-epoch relation for a concordance cosmology, with H$_0$ = 70 kms$^{-1}$Mpc$^{-1}$ Ω$_\Lambda$ = 0.75 and Ω$_M$ = 0.25. In comparing the model with observations, the same cosmology was used.

## 2. A TOY MODEL FOR GALAXIES AND THE POPULATION

The model that we have constructed to explore the issues that have been raised in the Introduction is intentionally very simple. It is intended to allow us to draw connections between phenomena in a semi-quantitative way, rather than to be a detailed physical model of the structural evolution of galaxies.

### 2.1 *New stars from star-formation*

Starting with small seed masses at very early epochs, the star-formation rate in a given galaxy is calculated using an evolving specific Star Formation Rate sSFR(*m,z*) that is appropriate for the majority of galaxies on the Main Sequence. We note that all masses and star formation rates in this paper will be expressed in terms of "long-lived" stars, enabling us to neglect the mass loss of a stellar population, i.e. they are the "reduced" quantities of Lilly et al (2013). One benefit of this is that galaxies that cease star-formation do not change their "stellar mass", easing comparisons of the galaxy at later epochs.

We adopt the following simple relation for the (reduced) sSFR

$$sSFR(m,z) = 0.07 \cdot \left(\frac{m}{3\times10^{10} M_\odot}\right)^{-0.2} \cdot (1+z)^2 \text{ Gyr}^{-1} \tag{1}$$

back to *z* = 6, but then have no further increase at higher redshifts. This change with redshift is broadly consistent with many observational estimates of this quantity (see Panella et al 2009, Stark et al 2014).

In a small increment of time, the mass of new (long-lived) stars formed in a galaxy of mass *m* is simply given by:

$$dm = m \cdot sSFR_{MS} \cdot dt \tag{2}$$

These new stars are distributed in an azimuthally symmetric exponential distribution with a scale length *h$_{SF}$* that is taken to be mass and redshift dependent. We adopt

$$h(m,z) = 5 \cdot \left(\frac{m}{3\times10^{10} M_\odot}\right)^{1/3} \cdot (1+z)^{-1} \text{ kpc.} \tag{3}$$

This incorporates the (1+*z*) scaling that was motivated in the Introduction. The *m*$^{1/3}$ scaling and the numeric pre-factor are chosen so that the model more or less reproduces the observed mass dependence of *R*$_e$(*m*) of star-forming galaxies at the present epoch (see Section 3.4). We note that the *m*$^{1/3}$ scaling mirrors that of dark matter haloes.

The increase in disk scale length of a given evolving galaxy over time arises from both the (1+z) and *m* terms in Equation (3), because the star-forming galaxy is increasing its mass as time passes. For a galaxy that stays on the Main Sequence, the former dominates at later epochs (z < 1) when not much mass is being added, while the latter dominates at earlier times (z > 2). Both result in inside-out growth of the given galaxy.

The increment $d\Sigma$ in surface mass density at a given radius *r* is then given by

$$d\Sigma(r) = \frac{1}{2\pi h_{SF}^2} \exp\left\{-\frac{r}{h_{SF}}\right\} \, dm \qquad (4)$$

The above relations for $h_{SF}$ and $sSFR_{MS}$ are mean relations. In order to produce a population of model galaxies with some scatter, individual galaxies are assigned $sSFR_{MS}$ and $h_{SF}$ that scatter around these mean relations by 0.02 and 0.2 dex respectively. For simplicity, this small offset is applied once and for all time. More complicated and no doubt more realistic schemes could have been implemented but would not be in the spirit of our toy-model.

A population of evolving galaxies is then constructed by following the evolution of one million galaxies whose distribution of initial seed masses, at redshift $z = 15$ is chosen from a power-law mass-function of logarithmic slope –0.4. This set of seed galaxies is designed to produce galaxies of relevant masses at the much later epochs that are of interest in this paper.

## 2.1 Quenching of star-formation

At any point in its evolution, the star-formation in a given model galaxy may be probabilistically quenched. Quenching results in the suppression of star-formation by a factor of 100 relative to Equation 1, effectively stopping the increase in stellar mass of the galaxy. For simplicity, quenching in this toy-model is assumed to be instantaneous, to occur simultaneously throughout the galaxy and to be a once-only process that is not reversible.

The probability that a galaxy is quenched within a given time step is taken from the mass-quenching formalism of Peng et al (2010). This probability can be simply expressed in terms of the mass increment *dm* in that time step that was given above in Equation 2 as

$$P \, dm = \frac{dm}{M^*} \qquad (5)$$

where M* is the (constant) characteristic mass of the Schechter function of the evolving star-forming galaxy population (see Peng et al 2010 for details). We stress that the probability to quench has no dependence whatsoever on size or mass-density.

While we are primarily concerned with the bulk of the galaxy population, i.e. the central galaxies, we will also wish to consider today's satellite galaxy population, to compare with the recent observational analysis of Omand et al (2014). To account for the additional quenching of these galaxies due to environmental effects we introduce, for the satellite population only, a further quenching probability term given by

$$P_{sat} \, dm = \epsilon_{sat} \frac{dm}{m} (1+z)^{-1} \qquad (6)$$

where $\epsilon_{sat}$ is the satellite quenching efficiency (van den Bosch 2008, Peng et al 2012). This satellite quenching has a probability that is only weakly dependent on stellar mass (because the logarithmic increase in density $dm/m$ varies only weakly with mass). As discussed above, this produces a second Schechter component to the mass function of passive galaxies whose faint end slope matches that of the star-forming population (see Peng et al 2012). The $(1+z)^{-1}$ dependence is introduced to approximately account for the fact that today's satellites will have become satellites over a wide range of epochs and that they will have been centrals, and thus immune to satellite-quenching effects, prior to becoming a satellite. It is not an important feature of the model and certainly does not produce the differences between centrals and satellites that are shown below, which have a quite different explanation.

For both types of quenching, we will assume that quenching occurs throughout the galaxy, simultaneously. We will also assume that there is no rearrangement of mass in the galaxy, i.e. that the surface mass density Σ(*r*) remains unaltered during and after quenching. Once stars are formed, they are assumed to remain at the same physical radius within the galaxy, i.e. there is no structural rearrangement of the stars in this toy model. Furthermore, no consideration is given as to the dimensionality of the structure (disk or spheroid) or to inclination effects. Each model galaxy is azimuthally symmetric.

## 2.3   *Calculation of mass and light profiles*

The final surface mass density profiles Σ(*r*) can therefore simply be integrated numerically by combining equations (1)-(4).

$$\Sigma(r) = \int d\Sigma(r) = \int \frac{1}{2\pi h_{SF}^2} \exp\left\{-\frac{r}{h_{SF}}\right\} \; m \cdot sSFR_{MS} \cdot dt \qquad (7)$$

The light profile μ(*r*) at any particular wavelength may then be estimated by the usual modification

$$\mu(r) = \int d\mu(r) = \int \frac{\Psi(t_0-t)}{2\pi h_{SF}^2} \exp\left\{-\frac{r}{h_{SF}}\right\} \; m \cdot sSFR_{MS} \cdot dt \qquad (8)$$

where Ψ(t') is the appropriate light-to-mass ratio of a coeval simple stellar population of age *t*', which in this paper we take from the Bruzual & Charlot (2003) models. We will only be interested in the light profile μ(*r*) in order to establish half-light radii (in the rest-frame *r*-band) of the galaxies, so the detailed choice of quantities like the initial mass function of stars should be unimportant.

## 2.4   *Mass functions and mass profiles*

In order to verify that this population of galaxies is reasonable in its non-structural properties, we look at the mass function of star-forming and quenched central galaxies at $z \sim 0$. This is shown in Figure 1 for both centrals and satellites. Not surprisingly, these are very similar to those expected from the Peng et al (2010) formalism. The central population consists of two Schechter functions describing the star-forming and passive populations. These have very similar M* but have α differing by Δα ~ (1+β) ~ 0.8 (see Peng et al 2010). The measured M* is close to the M* that was inserted as the quenching mass in Equation (5). In contrast, the faint end slope of the ϕ(*m*) of the passive satellites is essentially the same as the same as that the star-forming galaxies. We will see that this difference in the mass-functions of centrals and satellites plays an important role in the interpretation of the $f_Q(R_e,m)$ diagram.

Figure 2 shows the mean surface mass density profiles of star-forming galaxies at four different redshifts, choosing $10^{10.75}$ M$_\odot$ as an interesting representative mass scale (although any mass dependence will be modest, because of the weak dependence of sSFR on mass). In each case, the mass profile at large radii has an exponential profile. The mass profiles in the central parts rise well above this, due to the stars that formed at earlier epochs. If we operationally associate this inner excess with a "bulge" then B/T ratios of around 0.3 in light and 0.6 in mass are indicated for these galaxies (see Section 3.7 below). We would argue that these mass profiles are qualitatively not unreasonable for intermediate mass galaxies.

3. RESULTS

3.1 *Contributions to the apparent size evolution of active and passive galaxies*

Figure 3 shows the sizes of model star-forming galaxies, at fixed observed mass, as a function of redshift, choosing for this purpose the fiducial mass at M* ~ $10^{10.75}$ M$_\odot$. The points represent the half-light sizes of model star-forming galaxies and the solid blue line is the running mean of this population, here and elsewhere in the paper, averaging the log sizes. Not surprisingly, the half-light radius of the population of star-forming galaxies (at a given observed mass) evolves as $(1+z)^{-1}$. This simply reflects the input assumption of the model (Equation 3). The large points are data from Newman et al (2012) and serve to validate the adopted normalization of the $h_{SF}(m,z)$ relation.

The dashed blue curve shows the running mean of the *half-mass* sizes. The difference between the solid and dashed curves therefore represents the shrinking effect that would be observed when star-formation ceases due to the differential fading of the stellar population with radius, because of the implicit age gradient in this inside-out model. As noted above, this effect is driven by both $m$ and $(1+z)$ terms in Equation (3). The half-mass radii evolve as $(1+z)^{-0.85}$, i.e. a little slower than the half-light radii, because the shrinking effect due to fading increases with time, as discussed in Carollo et al (2014).

The solid and dashed red lines show the running mean log half-mass and half-light size of the *cumulative* population of quenched galaxies at the same fiducial mass of $10^{10.75}$ M$_\odot$. This population therefore contains all galaxies that have quenched at this particular mass at *any* earlier time. As expected, there is now very little difference between the half-light (solid) and half-mass radii (dashed) since light will closely follow mass in these passive galaxies. While very recently quenched galaxies will have the same half-mass radii as the star-forming galaxies at the same epoch, those that quenched earlier will have the half-mass radii of the star-forming galaxies (of the same mass) at the earlier epochs. The difference between the red line(s) and the blue dashed line therefore gives the effect of adding in all the previously quenched galaxies, i.e. the "progenitor effect". This will be driven by only the $(1+z)$ term in Equation (3). Because the passive galaxy population builds up cumulatively over cosmic time, the apparent size evolution of the passive population is always expected to be weaker than the apparent size evolution of the population selected to be star-forming. The mean log size of this quenched population in Figure 3 scales as $(1+z)^{0.6}$ (in both light and mass)

It can be seen that overall, the quenched galaxies have half-light radii that are typically about a half as large as the actively star-forming galaxies. This accords with what is seen observationally, e.g. in the relevant plots in Newman et al (2012) and van der Wel (2014). This difference in size is expected to increase slightly with epoch and is 0.4 dex at $z \sim 0$.

This factor of two change in half-light size from star-forming to quenched galaxies is, at all redshifts, roughly equally due to differential fading effects and to progenitor effects in the passive population. However, both of these two effects ultimately arise from the fact that star-forming galaxies were smaller in the past, through the $(1+z)$ and $m$ terms in Equation (3), as explained above.

As noted above, the evolution in the apparent size of the (cumulative) quenched population due to progenitor effects *alone* must always be less than the evolution in the star-forming objects. If the observed evolution in the quenched population is steeper than in the quenched population, then this would indicate the need for an additional effect (in addition to progenitor effects) that is not included in the toy model. This could be some redshift-dependent size change associated with quenching or a size increase after quenching, for example due to merging.

### 3.3  sSFR profiles and inside-out quenching

The central concentration of the early-formed stars causes a substantial radial gradient in the sSFR. At all redshifts, the model star-forming galaxies exhibit a pronounced radial gradient in sSFR. Figure 4 shows the average sSFR gradient in star-forming galaxies of mass $10^{10.75}$ $M_\odot$ at the four redshifts $z = 0,1,2,3$. The sSFR changes by an order of magnitude over the 1-10 kpc interval.

Such sSFR gradients of this strength have been observed in high redshift galaxies (Tachella et al 2015). As noted above, our quenching model suppresses star-formation throughout the entire galaxy simultaneously. Therefore the appearance of the pronounced sSFR gradient in the model galaxies is a direct consequence of the inside-out build-up of the stellar populations that is implied by the observed size evolution of the star-forming population. It need not have anything to do with a progressive radial onset of the quenching process.

### 3.4  Surface density thresholds and sSFR

We now turn to examine the key observational results that were highlighted in the Introduction, namely, the appearance of sharp surface density thresholds that appear to differentiate between the populations of actively star-forming and quenched, passively evolving, galaxies.

#### 3.4.1  The redshift dependent $\Sigma$-sSFR relation

In Figure 5, we show the mean log sSFR calculated for all galaxies $m > 10^8$ $M_\odot$ as a function of $\Sigma_{Re}$ at our four redshifts $z = 0,1,2,3$. It should be noted $\Sigma_{Re}$ is calculated both as the mean $\Sigma$ interior to the half-light radii (solid curves, as in Franx et al 2008 and most other observational studies) and interior to the half-mass radii (dashed curves). At each redshift, the curves from the model bear a striking resemblance to the data presented in Figure 9 in Franx et al (2008). At low redshift, the break in sSFR occurs at about $\Sigma_{Re} \sim 10^{8.5}$ $M_\odot$kpc$^{-2}$. More quantitatively, we can calculate a threshold $\Sigma_{Re}$ in exactly the same way as in Franx et al by computing the $\Sigma_{Re}$ at which the mean log sSFR has fallen by 0.3, i.e. the average sSFR has halved, or, crudely speaking, the quenched fraction $f_Q$ has risen to 50%. In Figure 6 we plot this threshold $\Sigma_{Re}$ against redshift to compare the model output with the observations of Franx et al. The quantitative agreement is astonishingly good, given the extreme simplicity of the toy model.

We conclude that the $\Sigma$-sSFR effects noted by Brinchmann et al (2004), Franx et al (2008), and by extension the low redshift $\Sigma$-D$_n$(4000) plot of Kaufmann et al (2003), may be mostly, or even completely, explained by the simple inside-out growth of galaxies that is *required* by the observed evolution in the $R_e(m,z)$ relation for star-forming galaxies *and have nothing to do with any physical link between surface mass density and the control of star-formation in galaxies*.

We note in passing that the change in mean log sSFR around the threshold $\Sigma$ is actually sharpest when the half-light radius is used to compute $\Sigma_{Re}$ rather than the half-mass radius. This extra effect can be simply understood in terms of the "fading effect" discussed in Section 3.1. This fading causes passive galaxies to be even smaller, in light, than their star-forming progenitors, and thus to have an even higher $\Sigma_{Re}$ (at a given $m$), further amplifying the difference in $\Sigma$ between star-forming and non-star-forming galaxies that comes from progenitor effects. The fact that the effect (in the model) is *stronger* using the light-defined radii than using the more physically meaningful mass-defined radii is a good example of the perils of using the relative tightness of correlations to imply an underlying causality.

#### 3.4.2  The local $f_Q(m,R_e)$ relation

Similarly, it is easy to construct from our model the quenched fraction of galaxies in the ($m,R_e$) plane, for comparison with the observed distributions from SDSS which were presented by Omand et al (2014). We do this for central galaxies in Figure 7 and for satellite galaxies in Figure

8. In both cases we again use the half-light $R_e$ for better comparison with the Omand et al analysis.

In both cases, the population produced by our model bears a striking resemblance to the data presented in the two panels of Figure 9 of Omand et al (2014). Not least, the imposing diagonal thresholds in $f_Q$ in their Figure 9 are reproduced in both Figure 7 and Figure 8. For the centrals (Figure 7) this threshold corresponds almost exactly to a locus of constant surface density. For the satellites, it is however noticeably shallower in the model population, as seen in the data of Omand et al (2014).

Both of these behaviors can be readily understood as follows. Almost all of the effects discussed in this paper stem from the fact that passive galaxies are, at all redshifts, smaller than the galaxies of the same stellar mass that are continuing to form stars. In our model, this is because they will have formed their stars at earlier epochs and higher redshifts, plus the differential radial fading of galaxies produced by age gradients implied by inside-out growth. As shown in Figure 3 in this paper (and as also observed in numerous studies) the typical size of this effect is about a factor of two when $R_e$ is estimated from the light distribution (0.4 dex at $z \sim 0$ in the specific implementation of Equation 3). With this established, we can them imagine two more or less parallel loci in the ($R_e$,$m$) plane, displaced by 0.4 dex in $R_e$, corresponding to the mean $R_e$-$m$ relations for the star-forming and passive galaxies. This is shown in Figure 9. These two parallel loci have slopes $R_e \propto m^{1/3}$, simply because of the size-mass relation adopted for newly formed stars (Equation 3).

For centrals, the different mass functions $\phi(m)$ that are produced for star-forming and quenched galaxies will populate these two loci differently (see Figure 1). In particular, the lowest mass galaxies will be almost entirely on the star-forming locus and the highest mass ones will be almost entirely on the passive locus. The ridge-line of the overall population (shown as the black line in Figure 9) will therefore shift from the star-forming locus at low masses to the passive locus at high redshift. This produces a *shallower* overall size-mass relation. Correspondingly, the contours of constant $f_Q$ will be *steeper* than each of the original two loci (and even more so when compared to the shallower overall ridge line).

It turns out that these contours of constant $f_Q$ are extremely well approximated by lines of constant $\Sigma_{Re}$, i.e. $R_e \propto m^{1/2}$ as seen in Figure 7. The fact that the iso-$f_Q$ contours are so closely parallel to lines of constant $\Sigma$ is in a sense a "coincidence" since it reflects the combination of the two more or less parallel $R_e \propto m^{1/3}$ loci (which comes from our input Equation 3) plus the relative numbers of galaxies along each, as given by the different shapes of the mass-functions of the two populations, which itself comes from the mass dependence of mass-quenching (Equation 5). It is this coincidence that makes the $\Sigma$ thresholds in the previously discussed plots from Kauffmann et al, Franx et al and Omand et al so impressively sharp.

Considering now the satellite galaxies, the passive mass function $\phi(m)$ has the same slope as the star-forming $\phi(m)$, especially at low mass where environmental-quenching dominates (Peng et al 2012). This is clearly seen in Figure 1. This means that the star-forming and passive loci (which are very similar to those for the centrals) are now more or less equally populated as a function of mass. As a result, the ridge-line of the overall population is now *parallel* to the two original loci, as will be the contours of constant $f_Q$. The reason why the iso-$f_Q$ contours are shallower for satellites than for centrals in Omand et al (2014) is therefore ultimately simply the different $\phi(m)$ for star-forming passive and star-forming galaxies that are produced by the different mass dependences of mass- and environment-quenching (Equations 5 and 6) and likely has nothing to do with any physical differences in the quenching or structure of satellites and centrals as suggested in that paper.

Our analysis has been based on a particular implementation of disk sizes, given by Equation (3), that includes both a mass and redshift dependence, and one might worry that the results are dependent on the precise exponents of these two dependencies. Recall however that the final iso-$f_Q$ contours are steepened from the underlying size-mass relations for star-forming and passive galaxies because of the differential populating of the mass-functions, coupled with the size offset between the two size-mass relations that itself comes (Figure 3) from the fading and progenitor effects produced by the $m$ and $(1+z)$ terms. The differential populating, which come from the different $\alpha$ and the attendant increase of $f_Q$ with mass, is impossible to avoid, at least for centrals. Therefore, any size-mass relation for star-forming galaxies that is a bit shallower than $r \propto m^{1/2}$, e.g. the $r \propto m^{1/3}$ adopted here (based on Omand et al 2014 Figure 3) can be easily steepened up via the size offset between star-forming and passive galaxies to have the iso-$f_Q$ contours closely parallel to lines of constant $\Sigma$. This size offset will be more or less proportional to the strength of the $(1+z)$ dependence in Equation (3) but, perhaps counterintuitively, a smaller offset can produce more steepening (indeed having no offset at all will produce vertical iso-$f_Q$ contours). As an example, substituting a weaker $(1+z)^{-1/2}$ evolution in Equation (3) produces a smaller offset of about 0.25 dex at $z \sim 0$, but this produces slightly steeper iso-$f_Q$ contours, parallel to $r \propto m^{0.58}$. These however still quite close to the line of constant $\Sigma$. We conclude that the basic arguments advanced in this paper do not sensitively dependent of the precise exponents of $m$ and $(1+z)$ in Equation (3).

One final point is quite interesting. The basic model described above fails to reproduce the curved upturn in the quenched fraction above $m \sim 10^{11}$ M$_\odot$ of Omand et al's Figure 9, i.e. it does not appear to produce enough quenched galaxies at very high masses with low surface density. We know however from the continuity arguments of Peng et al (2010) that these highest mass galaxies at $m > 10^{11}$ M$_\odot$ are the ones that are most likely to have undergone significant merging after quenching. This mass threshold is clearly associated with other signatures of significant merging in the passive population, including boxy isophotes (Bender et al 1988), metallicity gradients (Carollo et al 1993), cores (Faber et al 1997), and v/σ signatures (Rix et al 1999). It is therefore quite likely that merging has modified the structures of these most massive passive galaxies. To test this idea, we can add (for this part of the paper only) a very crude representation of the effects of homologous merging by increasing the masses and radii of all passive galaxies by 0.2 dex, thereby decreasing all the $\Sigma_e$ by 0.2 dex. This has the effect of shifting these galaxies diagonally in the $m$-$R_e$ plane. As expected this has little effect at low masses, but has the desired result at masses above $10^{11}$ M$_\odot$ as shown in Figure 10. Correspondingly, merging is likely the cause of the upwards curvature to the size-mass relation for passive galaxies that occurs above $10^{11}$ M$_\odot$ in the lower panel of Figure 3 in Omand et al (2014).

The point of this discussion is to show that the basic form of the SDSS $f_Q(m,R_e)$ as presented by Omand et al (2014) can be fully understood in terms of the (observed) offsets between the size-mass relations for star-forming and passive galaxies. Furthermore, we argue that this offset follows very naturally from the (observed) size evolution in the population of star-forming galaxies, through a combination of fading and progenitor effects that themselves follow from the $m$ and $(1+z)$ scalings in Equation (3). We stress that Figures 7 and 8 are generated from a model in in which there is no dependence of quenching on size or density. The sharp apparent thresholds in $\Sigma$ in Omand et al (2014) and in Brinchmann et al (2004) and Franx et al (2008) need therefore contain no information at all on the quenching mechanism(s). Not least, Omand et al's argument (their Figure 15) that the iso-$f_Q$ contours in this plane should be vertical in any model with purely mass-dependent quenching (as in e.g. Peng et al 2010) is clearly not correct once the size growth of the progenitor population of star-forming galaxies is taken into account.

### 3.5 *Surface mass density evolution of individual galaxies*

We can also look at the evolution of individual galaxies as they build-up their stellar mass. Figure 11 shows the evolution of the sSFR and central mass density $\Sigma_{1kpc}$ for 20 representative star-forming galaxies of $10^{10.75}$ M$_\odot$ that have survived to the present-day (tracks in blue), plus 20 representative passive galaxies of the same present-day mass that however quenched at earlier times (tracks in red). The tracks of the passive galaxies give the appearance of growing to some threshold $\Sigma_{1kpc}$ and then quenching around this critical surface density. This impression is reinforced by the fact that the surviving star-forming galaxies have generally not reached this same $\Sigma$. However, as elsewhere in this paper, this apparent role of $\Sigma$ in the quenching of star-formation has no (direct) physical basis: it is a consequence of the link between stellar density and epoch implicit in the size evolution of star-forming galaxies.

### 3.6 *Differential growth of passive galaxies*

As an aside, Figure 12 shows the evolution with redshift of the average surface mass densities of the passive galaxy population at $10^{10.75}$ M$_\odot$ that is calculated within the inner 1 kpc, $\Sigma_{1kpc}$, and within the half-light $R_e$, $\Sigma_{Re}$. The evolution with redshift in these quantities is due entirely to the progenitor effects discussed in the previous Section. It can be seen that the apparent evolution in $\Sigma_{Re}$ of the passive galaxies is significantly larger than the evolution in $\Sigma_{1kpc}$. There is some evidence for this observationally (van Dokkum et al 2010). This differential effect could be interpreted as indicating that the galaxies are puffing up by preferentially adding mass and kinetic energy in their outer parts through minor mergers. Our toy model indicates that some or even all of a differential effect of passive size growth with radius can be accounted for by the progenitor effects within the passive population that ultimately follow from the size evolution of star-forming galaxies.

### 3.7 *Mass-dependent B/T ratios*

Using our mass and light profiles (e.g. from Figure 2) we can estimate a B/T ratio. We here adopt a heuristic approach and empirically define the "bulge" to be the excess of mass, or light, that lies above the inwards extrapolation of the outer exponential profile. As shown in Figure 2, the mass and light profiles produced by the toy model all have "bulges", defined in this way. Of course there are no kinematics, or 3-dimensional structures, in this model and so the association with real galactic bulges is not intended to be exact.

The "bulge" defined in this way will consist of older stars formed in smaller disks at earlier times and the mass-profile will reflect the star-formation history of the galaxy. Once a galaxy quenches, the fading of the younger disk will further enhance the appearance of the older central bulge.

In the paper we have so far focussed on a single fiducial mass of $10^{10.75}$ M$_\odot$. This was because differences in the mass (or light) profiles of galaxies in the model can only arise from differences in the relative histories of stellar mass production. The fact that the sSFR in the toy model varies only rather slowly with mass ($\beta \sim -0.2$ in Equation 1) means that all galaxies that are still star-forming will have had rather similar star-formation histories and therefore will have similarly shaped profiles. Generally, galaxies that have a higher sSFR will have grown more rapidly and have a smaller fraction of older stars in high-density components, leading to the appearance of smaller bulges.

The apparent B/T ratios in the toy model for star-forming and quenched galaxies are shown at two redshifts, z = 2 and z = 0, in the left hand panels of Figure 13. The apparent B/T ratios of star-forming galaxies are systematically lower than the B/T for the quenched population, reflecting the much lower mass-to-light in the young star-forming disks of the latter. There is a trend towards higher B/T at higher masses, but it is quite weak, reflecting the small negative value of $\beta$.

To explore further the link between the profiles of galaxies and their star-formation histories, we modify the Main Sequence sSFR-mass relation (Equation 1) to suppress the sSFR at high masses. We introduce a curvature into the Main Sequence by suppressing the sSFR above $10^{10}$ M$_\odot$ by a mass-dependent factor $f(m)$ that is given by

$$\log f(m) = -0.3 \, (\log m_{10})^2$$

where $m_{10}$ is the stellar mass in units of $10^{10}$ M$_\odot$. There is some evidence for such a flattening of the increase of SFR along the Main Sequence (see Schreiber et al 2015). This modified sSFR-mass relation is shown in the upper right panel of Figure 13, compared with the original Equation 1 in the upper left panel. This modification has the expected effect on the B/T values. At high masses, the galaxies have a lower relative sSFR and so will have formed relatively more of their stars at earlier times. They will therefore have, in our toy model, a more prominent "bulge".

The upwards curvature of the B/T ratio on the left side of Figure 13 is therefore directly related to the downwards curvature of the Main Sequence, and starts at the same mass. In our toy model, the former follows from the latter, whereas an observer, confronted with these data, might be tempted to conclude that the sSFR had in fact been suppressed in some way *by the presence of the bulge*. Of course, we would still have to explain physically why the sSFR of high mass Main Sequence galaxies was suppressed. However, there are several possibilities for this that could be completely unrelated to the galactic internal structure. The point of this discussion is simply to emphasize, again, the difficulty of establishing cause and effect in the correlations between the gross properties of galaxies, especially if other known effects, in this case the evolution of the size-mass relation for star-forming galaxies, are not fully taken into account.

4.  DISCUSSION

Our paper has aimed to strike a cautionary note in the interpretation of observational data on the galaxy population. In particular, the above discussion has highlighted the dangers of inferring causality from even very tight observational correlations unless the consequences of all other properties of the galaxy population are properly followed.

We have argued that a purely mass-dependent quenching law of the form advanced by Peng et al (2010) inevitably produces a strong apparent dependence of the quenched fraction (or mean log sSFR) on surface mass density $\Sigma$, once the observed evolution of the size-mass relation for star-forming galaxies is taken into account. Our toy-model produces a very sharp apparent threshold in $\Sigma$ between star-forming and quenched galaxies. By the same token, however, it is always possible that the real physical situation is the reverse. It could be that the physical quenching behavior is indeed driven by the surface mass density and that the simple mass-dependent quenching laws of Peng et al (2010) are themselves the side-effect *consequence* of this plus the size evolution: we could not tell the difference from analyses of the $f_Q(m,R_e)$ or $f_Q(m,\Sigma)$ surfaces.

We have focused on the surface mass-density $\Sigma$ as a convenient parameterization of structure because it is simple to calculate. We have ignored both the kinematic properties and the dimensionality of galactic structures, i.e. the 2-dimensionality of disks and the 3-dimensionality of spheroids. The kinematics will, like $\Sigma$, follow from a combination of $m$ and $R_e$, from virial arguments, and so the arguments in this paper should apply also. The dense central parts of galaxies, which we have operationally called bulges in Section 3.7, are generally spheroidal in real galaxies. Except at the highest masses above $10^{11}$, these spheroids will generally be rapidly rotating and we would simply need a way to ensure that stars that form in dense (disk-like) structures at early times end up, at later epochs, in rotating spheroids. The familiar processes of heating through merging or disk instabilities are likely responsible for this (see Noguchi 1999, Navarro & Steinmetz 2002, Kormendy & Kennicutt 2004, Dekel et al 2009 amongst many others). As a first approximation, any mechanism that spheroidizes most of the stars formed at high redshifts, e.g. $z > 1.5$-$2.0$ would work.

The reader is referred to discussions along similar lines in Driver et al (2013). The difference with that work is that we would have high redshift stars made in disks and ending up in spheroids, whereas Driver et al envisage a different mode of star-formation at high redshifts that produces stars in spheroids, but the outcome is similar.

So, are their any observational tests of our proposed scenario? A clear one is that within the quenched population, the "age" of the stellar population (specifically the elapsed time since quenching) at a given stellar mass should correlate with the surface density, or inversely with the size (see Fagioli et al, submitted). However, tests of this prediction will need to carefully consider the mass range selected. As discussed in Section 3.4.2, above $10^{11}M_\odot$, the fraction of galaxies that have undergone merging will increase significantly, and this may well lower the average surface density. In general, we would expect that the galaxies that have merged most, i.e. which were today the most massive, would have quenched first, i.e. they would have reached the quenching mass earliest to maximize the time for merging. This could well eliminate or even reverse the correlation between age and $\Sigma$, especially at high stellar masses.

A more general caution involves interpreting the tightness of correlations or the sharpness of thresholds as being indicative of causality. We have emphasized this in terms of the tighter observed correlation between quenching and surface mass density than between quenching and mass. However, the $\Sigma_e$ threshold between star-forming and quenched galaxies (e.g. in Figure 5) is clearly sharper when the $\Sigma_e$ is calculated using half-light radii than when using half-mass radii. This is because of the added effect of differential fading of the outer parts of the galaxies and clearly does not imply anything about the relative physical importance of half-light radii and half-mass radii – indeed our analysis has suggested that neither may be playing any role in quenching at all!

Finally, we comment that the analysis presented here highlights the difficulties of inferring the physical characteristics of quenching by comparing star-forming and quenched galaxies at a single epoch (whether at high redshift or locally), without taking into account the obvious fact that the quenched population will be dominated by galaxies that actually quenched a long time previously, when the star-forming population was in all likelihood quite different. This is especially true when considering parameters such as size, metallicity, or halo mass that are either observed, or would be expected theoretically, to evolve with redshift for either or both of the star-forming or quenched populations.

5. SUMMARY

The goal of this paper has been to demonstrate that a number of different structure-related effects that are seen in the galaxy population can be straightforwardly understood to be a consequence of the observational fact that the half-light radius of star-forming galaxies at fixed mass increases with time. This observed size-evolution creates a broad link between the density of a stellar population and when it forms, or equivalently between the overall density and the star-formation history of a given galaxy. This connection underlies most of the other effects of interest.

This has been demonstrated via a simple toy-model of growing galaxies in which the new stellar mass from star-formation is added to galaxies in exponential components whose scale length scales as $m^{1/3}$ and $(1+z)^{-1}$. The mass to be added is determined from the SFR via the overall evolution of the sSFR with cosmic time. Galaxies quench their star-formation according to empirical probabilistic laws that depend only on the total mass of the galaxy and not at all on the surface mass density or size. Furthermore quenching occurs instantaneously throughout the galaxy. For simplicity and transparency, we assume (except for the brief digression in Section 3.4.2) that there is no re-arrangement at all of stellar mass during or after the quenching process, i.e. that the projected mass profile $\Sigma(r)$ remains unchanged during and after quenching. When necessary, we consider satellite galaxies separately by introducing an additional probability of environment-quenching, but the structural outcome of both quenching channels is assumed to be identical.

This very simple toy-model reproduces several inter-related features of the evolving galaxy population that appear to involve size and/or surface densities. Specifically, the model reproduces the following observational facts and correlations:

1. The population of passive galaxies have half-light sizes that are a factor of about two smaller at a given mass and epoch than the star-forming ones, despite the fact that there is no change in the mass profile of galaxies during or after quenching in the model. Roughly 50% of this (apparent) shrinking effect is due to differential fading with radius, because of the underlying age gradient in the star-forming progenitors, and 50% is due to progenitor effects that arise from the broad range of quenching epochs for the passive galaxies, and the consequent cumulative build-up of the population over time and the inclusion of compact galaxies that quenched at early times.

2. There is a sharp apparent threshold in surface mass density between actively star-forming and passive galaxies, even though the empirical quenching laws used in the model depend only on the integrated stellar mass, and not at all on the stellar density. The mean log sSFR changes abruptly at a threshold $\Sigma_e$, as in Brinchmann et al (2004) and Franx et al (2008). Furthermore, this threshold $\Sigma_e$ increases with redshift back to $z \sim 3$ exactly as observed in the latter work.

3. The half-light radius $R_e$ appears as a strong parameter in the fraction of galaxies that are quenched at a given mass. The overall variation of quenched fraction with radius and mass $f_Q(R_e,m)$ of SDSS centrals and satellites presented by Omand et al (2014) is well reproduced. The appearance of a striking diagonal demarcation between star-forming and quenched galaxies that is parallel to the line of constant $\Sigma_e$ (equivalent to the sharp threshold $\Sigma$ in the previous point) is completely explained in our toy-model without any size or density dependence of the quenching probabilities. The fact that the iso-$f_Q$ contours are exactly parallel to the lines of constant surface mass density for central galaxies is shown to arise from the combination of the slopes of the underlying size-mass relation for star-forming and passive galaxies, the offset between them, and the relative mass-functions of these populations and is in a sense a coincidence.

4. The different slope of the iso-quenched-fraction contours of satellites compared with centrals is also easily explained simply in terms of the different mass functions of the two populations and, again, likely contains no physical information about any differences in the quenching of satellites and centrals, since the model treats these identically.

5. The one feature of the $f_Q(R_e,m)$ distribution that is not explained by our basic model, which is the presence of low $\Sigma$ quenched galaxies of very high mass, likely reflects the effects of post-quenching merging of galaxies at these very highest masses.

6. The model exhibits radial gradients in sSFR because of the presence of older denser populations in the centers of the galaxies. This should not be thought of as being caused by a progressive "inside-out" quenching process but rather simply reflects the differential build-up of stellar mass implied by the observed evolution of the size mass relation of star-forming galaxies.

7. The model also produces an apparent inside-out growth of galaxies in the passive population in the sense that the mean mass density within 1 kpc changes less with epoch than that within $R_e$. This arises in our model purely from progenitor effects rather than any differential addition of mass to quenched galaxies, since no mass is added after quenching.

The analysis mostly focused on a single fiducial mass around M*. Differences in the mass-density profiles of galaxies with mass can arise (in the model) from differences in the SFR histories of the galaxies. We show the connections between the light profiles, crudely parameterized as a bulge-to-total ratio, and the star-formation histories. Our standard model, in which the sSFR is only a weak power of stellar mass, produces B/T ratios that increase with mass, but only weakly. Suppressing star-formation in high mass Main Sequence galaxies by putting in a downwards curvature to the Main Sequence at high masses produces a corresponding upwards curvature in the apparent B/T ratios with mass. Although there is in this sense a connection, our toy-model shows that there is no need to have a causal link between the presence of the bulge and the suppression of star-formation.

As with all such toy-models, the point of this analysis has not been to present a detailed physical model for galaxy evolution. Rather it has been to show that, starting only from the observed evolution in the size-mass relation for star-forming galaxies, a number of different observational effects would be expected that could easily be misinterpreted as reflecting some underlying physical causality when, in fact there may be no causal connection at all. Interpretation of such observed correlations should therefore be approached with appropriate caution.

The analysis presented here also highlights the dangers of inferring relative importance from the relative tightness of observational relations, and the hazards of trying to infer the physical characteristics of quenching by comparing star-forming and quenched galaxies at a single epoch, without properly taking into account the fact that the quenched population will be dominated by galaxies that quenched long ago when the star-forming population was likely to have been quite different.

Acknowledgements:   This work has been supported by the Swiss National Science Foundation.

*Figure 1: The present-day mass function of star-forming and passive central (left) and satellits (right) galaxies at the present epoch in the model population. The overall vertical normalisation is arbitrary. As expected in the Peng formalism, the two central populations have the same M\* but faint end slopes α that differ by about 0.8 ~ (1+β), while the two satellite populations have the same faint end slope. This difference is responsible for some of the effects discussed in this paper (see Figure 9). The thin lines show a single Schechter function fitted to each population.*

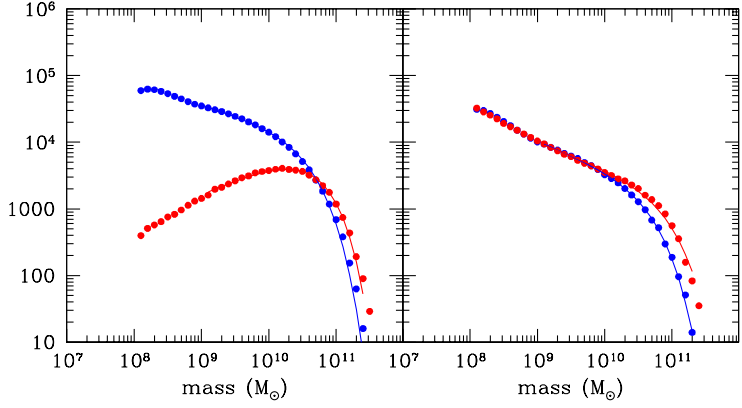

*Figure 2: The mean radial mass profile for star-forming galaxies of observed mass $10^{10.75}$ $M_\odot$ at different redshifts: z = 3 (black), z = 2 (blue), z = 1 (magenta) and z = 0 (red).*

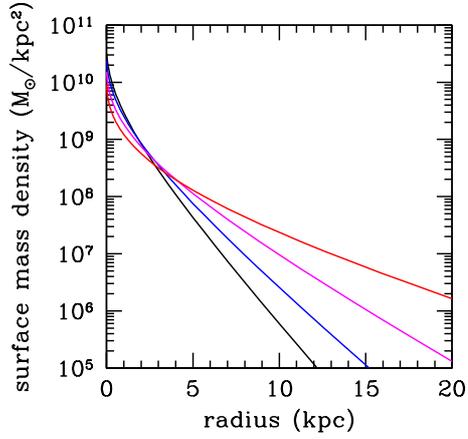

*Figure 3: The sizes of model galaxies as a function of redshift at a fiducial mass of $10^{10.75}$ $M_\odot$. The small black dots show the $R_e$ half-light radii (in the R-band) of individual star-forming galaxies. The solid blue line is the mean of this population as a function of redshift and should be compared with the large black dots showing observational data from Newman et al (2012). The good agreement justifies the choice of $h_{SF}(m,z)$ in the toy-model. The dashed blue curve then shows the mean of the half-mass sizes, which are systematically smaller because of the radial variation in mass-to-light ratio. The difference between the two blue lines therefore gives the shrinking in apparent size that would be expected when a star-forming galaxy quenches without rearrangement of mass. The red lines then show the running mean of the half-light and half-mass radii for the population of quenched galaxies. The difference between the red line(s) and the blue dashed line therefore gives the effect of adding in the previously quenched galaxies, i.e. the "progenitor effect". At all redshifts, the quenched galaxies are about a factor of two smaller in light, as observed. In the model, about a half of this comes from differential fading and a half from the progenitor effect.*

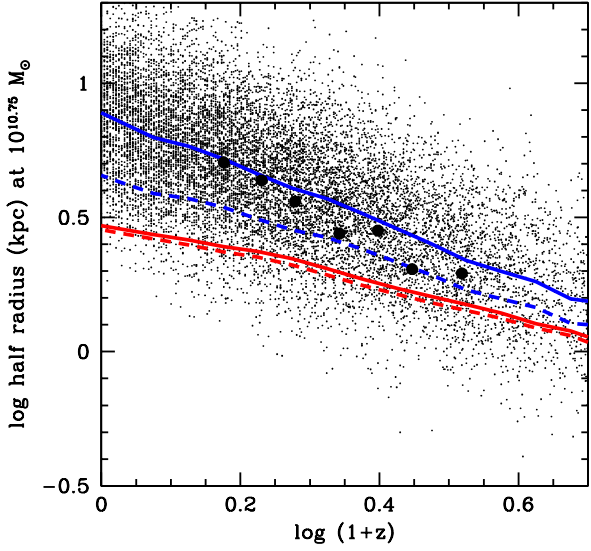

*Figure 4: The radial variation of average sSFR in star-forming galaxies of mass $10^{10.75}$ $M_\odot$ at the four redshifts z = 0,1,2,3 (colour-coded as in Figure 2), The sSFR changes by an order of magnitude over the 1-10 kpc interval. This is a consequence of the inside-out growth of galaxies in the model and is not due to any differential onset of quenching at different radii or surface densities.*

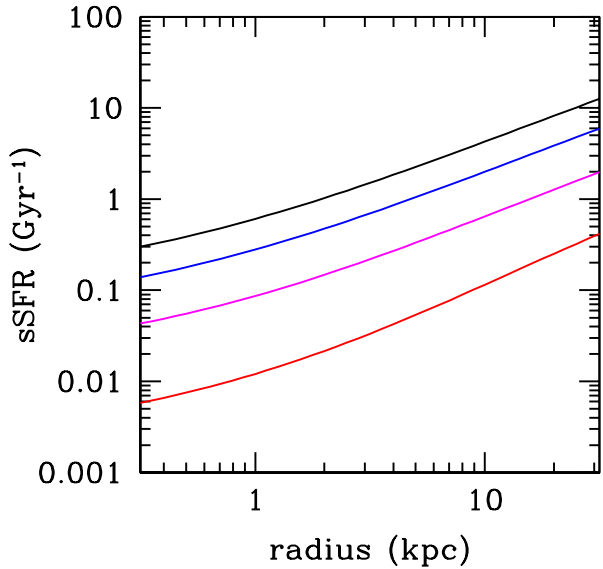

*Figure 5: The mean log sSFR calculated for galaxies (m > 10⁸ $M_\odot$) as a function of the mean surface density within the half-light radius, as in Franx et al 2007 (solid curves) and within the half-mass radius (dashed curves). There is a sharp break at a threshold mass density. This threshold mass density increases with redshift, as observed. Note that the break appears to be sharper when the half-light radius is used. This is because of the extra shrinking of galaxies as their outer parts fade, shown on Figure 3.*

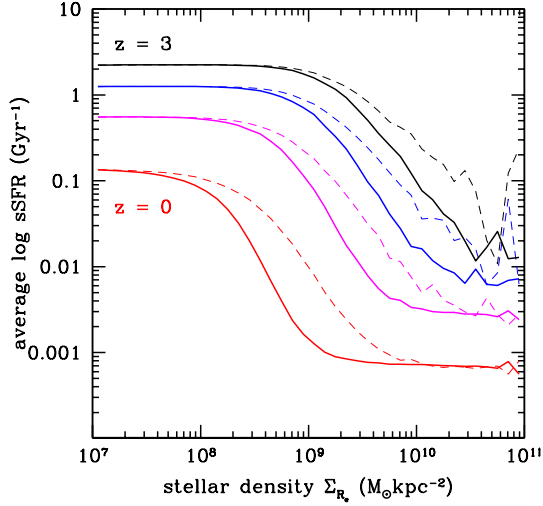

*Figure 6: The change in the threshold surface density (solid dots), computed as in Franx et al (2007), compared with the observed values from that paper (stars). The variation with redshift is in remarkably good agreement. The open dots are the thresholds computed using the half-mass radii.*

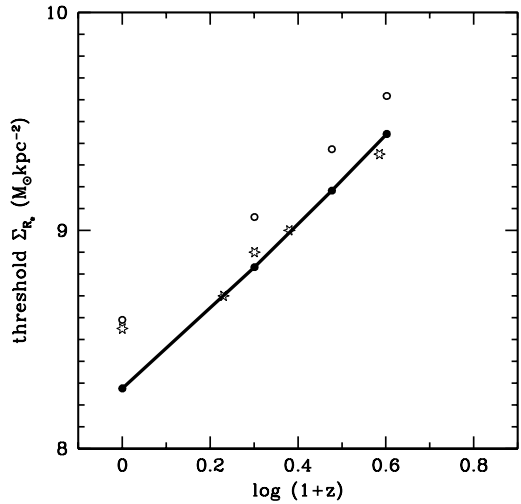

*Figure 7: The variation of the quenched fraction $f_Q(m,R_e)$ in the model for central galaxies, for comparison with the SDSS data presented by Omand et al (2014) in their Figure 9. The model successfully reproduces the strong diagonal threshold in $f_Q$ that corresponds almost exactly to a line of constant surface mass density. This demonstrates that this striking feature can be produced in a model in which quenching depends purely on mass and not in any way of surface mass density. Why the threshold is so nearly a line of constant surface mass density is explained in the text and by Figure 9. The sharpness of this transition is why the step function is Figure 5 is so sharp, and also why Figure 5 does not depend on the adopted mass range.*

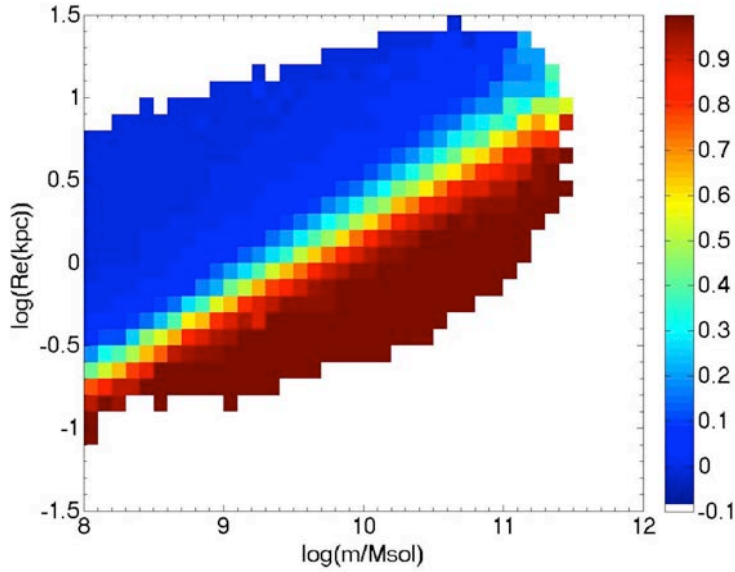

*Figure 8: As for Figure 7, but for satellite galaxies. The demarcation between star-forming and quenched galaxies is shallower, as seen in the data of Omand et al (2014), despite the fact that in the model there is no difference in the quenching of satellites and centrals beyond the differences in mass-dependence. This is readily explained by the different mass-functions of centrals and satellite galaxies - see text and Figure 9.*

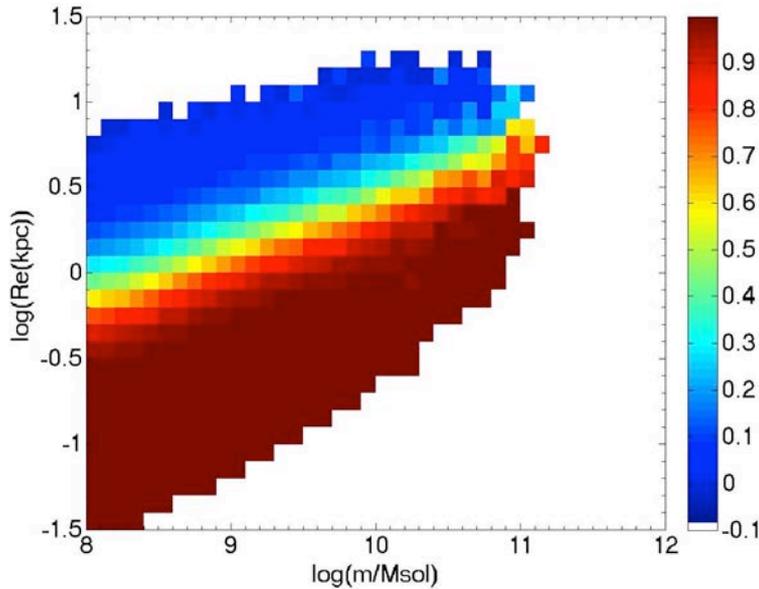

*Figure 9: This diagram explains the origin of the strong Σ demarcation between star-forming and central galaxies in $f_Q(m,R_e)$ and the origin of the difference between centrals (upper panel) and satellites (lower panel) as seen on Figures 7 and 8 respectively. The blue and red solid lines show the mean relations for star-forming and passive galaxies. These are very similar in both panels and roughly follow the $R_e \propto m^{\frac{1}{3}}$ size-mass relation that was input to the model. The black line gives the mean relation for the overall population, which differs in the two panels because of the pronounced difference in the mass functions of star-forming and passive galaxies for centrals and satellites shown on Figure 1. The light magenta contours give the corresponding iso-$f_Q$ contours, which are steeper for centrals than for satellites because of the shallower black line that results from the different mass-functions along each locus. See text for further explanation.*

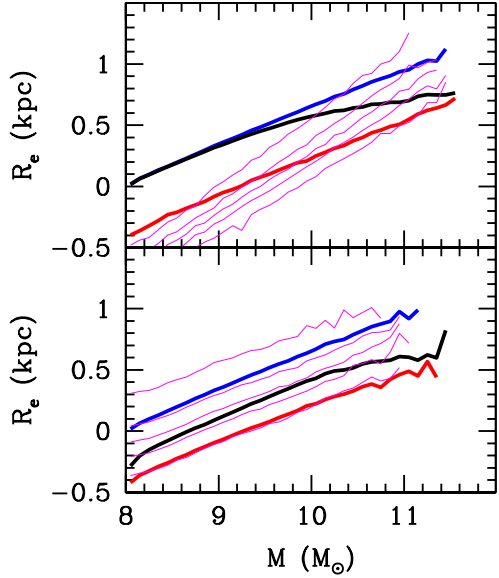

*Figure 10: As for Figure 7, but incorporating a simple representation of the effect of a modest amount of homologous merging of passive galaxies post-quenching. This successfully reproduces the appearance of passive galaxies with low surface densities at very high masses and the upturn in the demarcation line at high masses.*

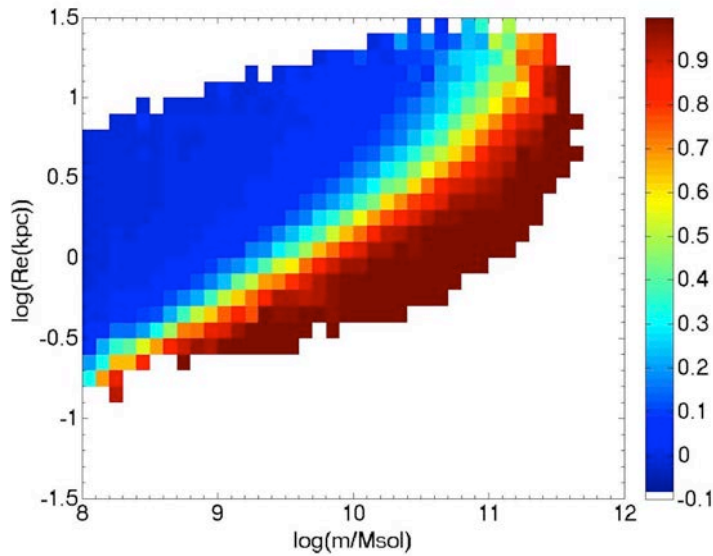

*Figure 11: The evolution of overall sSFR and central mass density $\Sigma_{1kpc}$ for 20 representative star-forming galaxies that end up today with mass $10^{10.75} M_\odot$, and for 20 representative passive galaxies that quenched at this mass at some earlier time. The appearance of a surface mass density threshold, beyond which galaxies quench, has no physical basis in the model.*

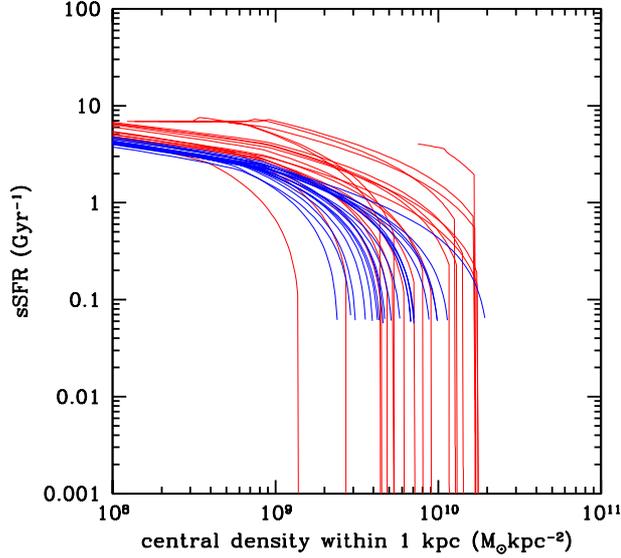

*Figure 12: The average surface mass densities of the passive galaxy population at $10^{10.75} M_\odot$ that is calculated within the inner 1 kpc, $\Sigma_{1kpc}$, and within the half-light $R_e$, $\Sigma_{Re}$. The evolution with redshift in these quantities is due entirely to the progenitor effects. The evolution is significantly shallower in the central density, as observed, but this is not due to any differential puffing up of the galaxies due to the addition of mass or kinetic energy to the galaxies, since no mass is added following quenching.*

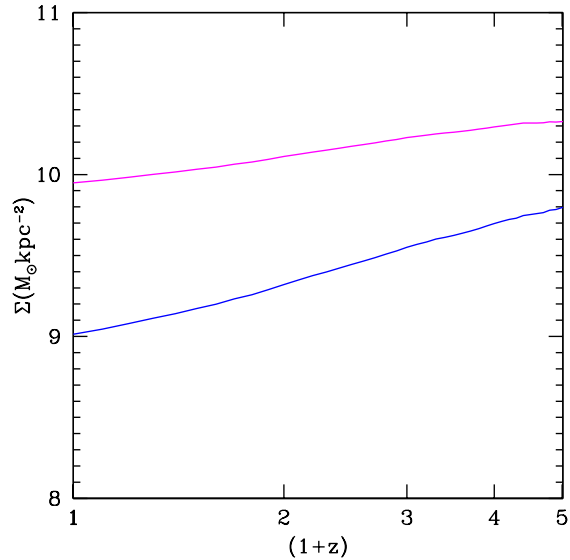

*Figure 13: The apparent B/T ratio, as defined in the text, for toy model galaxies that are star-forming (blue) and quenched (red) at z = 2 (middle panels) and z = 0 (lower panels) over a range of masses. The panels on the left are produced by the standard main sequence sSFR(m,z) given by Equation 1 and plotted in the upper left panel. The systematic difference between red and blue points is primarily due to the low mass-to-light ratios of the large star-forming disks in the latter. Both blue and red sequences have only a gradual change in profile with mass, because of the weak dependence of SFR history on mass produced by the small value of β. However, the introduction of curvature into the Main Sequence sSFR at high masses, as shown in the upper right panel, produces a corresponding curvature in B/T with mass, simply due to the star-formation histories rather than any physical link between the presence of a bulge and the suppression of sSFR in the galaxy.*

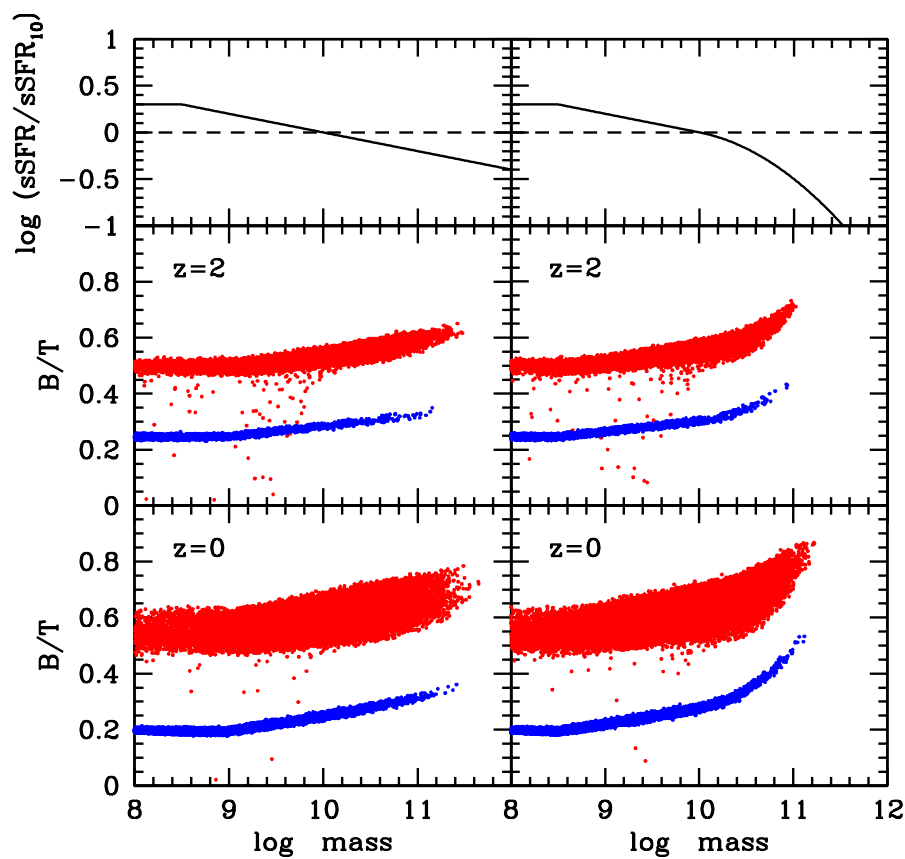